\documentclass[
  ,draft            % use draft while you are working on the paper
  ]
  {aipproc}

\layoutstyle{6x9}

\def\lapp{\ifmmode\stackrel{<}{_{\sim}}\else$\stackrel{<}{_{\sim}}$\fi}

\begin{document}

\title{Pulsar Timing Arrays and their Applications}


\author{R. N. Manchester}{
  address={CSIRO Astronomy and Space Science, Australia Telescope
    National Facility,\\ PO Box 76, Epping NSW 1710, Australia}
}

\begin{abstract}
  Millisecond pulsars are intrinsically very stable clocks and precise
  measurement of their observed pulse periods can be used to study a
  wide variety of astrophysical phenomena. In particular, observations
  of a large sample of millisecond pulsars at regular intervals,
  constituting a Pulsar Timing Array (PTA), can be used as a detector of
  low-frequency gravitational waves and to establish a standard of
  time independent of terrestrial atomic timescales. Three major
  timing array projects have been established: The European Pulsar
  Timing Array (EPTA), the North American pulsar timing array
  (NANOGrav) and the Parkes Pulsar Timing Array (PPTA). Results from
  the PPTA project are described in some detail and future prospects
  for PTA projects are discussed.
\end{abstract}

\maketitle

%%%%%%%%%%%%%%%%%%%%%%%%%%%%%%%%%%%%%%%%%%%%
%% MAINMATTER
%%%%%%%%%%%%%%%%%%%%%%%%%%%%%%%%%%%%%%%%%%%%

\section{Introduction}
Pulsars and especially millisecond pulsars (MSPs) are remarkably
stable celestial clocks. This great period stability opens up a wide
range of potential applications. Pulsar timing analysis is based on
the measurement of precise pulse times of arrival (ToAs) at the
telescope. These ToAs are then transformed to the Solar-System
barycentre which approximates an inertial frame. A model for the
pulsar, including its position, proper motion, period and period
derivatives and binary parameters (if appropriate) can be used to
predict the pulse ToAs. The difference between the observed and
predicted ToAs are known as {\em timing residuals}. These timing
residuals contain information about errors in the model parameters and
unmodelled phenomena affecting the observed pulse period and so are at
the heart of all pulsar timing analyses; see Hobbs et al. \cite{hem06}
and Edwards et al. \cite{ehm06} for more details.

A pulsar timing array (PTA) consists of an array of pulsars, widely
distributed on the celestial sphere, that are being timed with high
precision and at frequent intervals over a long data span
\cite{hd83,rom89,fb90}. Such a PTA has the potential to detect
low-frequency gravitational waves propagating in the Galaxy
\cite{saz78,det79}, to improve our knowledge of Solar-System
parameters \cite{fb90} and to establish a pulsar-based standard of
time that is independent of terrestrial atomic timescales
\cite{pt96}. Only MSPs have sufficiently narrow pulses (in time units)
and sufficiently stable periodicities to be useful for PTA
applications.

\section{Detection of Gravitational Waves}
Gravitational waves (GWs), a prediction of Einstein's general theory
of relativity, are fluctuations in spacetime generated by acceleration
of massive objects which propagate at the velocity of light. Possible
astrophysical sources include energy-density fluctuations in the
inflation era, oscillations of cosmic strings, binary supermassive
black holes in the cores of distant galaxies and double-neutron-star
binary systems. Although the famous Hulse-Taylor binary pulsar has
given the first observational evidence for the existence of
gravitational waves through its orbital decay \cite{tfm79,wt05},
gravitational waves have never been directly detected despite huge
efforts over more than four decades. Current projects such as {\em
  LIGO} \cite{aaa+09j} and {\em Virgo} \cite{aaa+08a} utilise
laser-interferometer systems that are sensitive to GWs with
frequencies in the range 40 -- 1000~Hz. The planned space
gravitational-wave observatory {\em LISA} \cite{sha08} has much longer
interferometer arms and is sensitive to GWs in the frequency range 0.1
-- 100~mHz. In contrast, PTA systems are most sensitive to signals
with frequencies about $1/T$, where $T$ is the data span of the timing
observations. Typically, $T\sim 10$~yr, corresponding to frequencies
of a few nHz.

In this nHz band, the strongest GW signal is expected to be a stochastic
background from binary super-massive black holes in the cores of
distant galaxies \cite{jb03,wl03a,svc08}. Detection of this expected
signal requires high-quality timing observations of about 20 MSPs over a
data span of at least five years with timing precisions of the order
of 100 ns \cite{jhkm05}. This level of timing precision can only be
obtained for a few MSPs with present technology but fortunately the expected
signal has a very ``red'' spectrum (stronger at low frequencies) and
so the required sensitivity can be recovered by observing for longer
data spans. Specifically, for the stochastic background from binary
super-massive black holes, the expected signal may be represented by 
\begin{equation}
h_c(f) = A (f/f_{\rm 1 yr})^{\alpha}
\end{equation}
where $h_c(f)$ is the characteristic GW strain at frequency $f$, $A$
is the dimensionless amplitude of the GW background, $f_{\rm 1 yr} =
({\rm 1 yr})^{-1}$ and the spectral index $\alpha = -2/3$
\cite{jhkm05}. This leads to a $T^{13/3}$ dependence for the
amplitude of the signal in the timing residuals \cite{hjl+09}.

The absence of any unmodelled signal (apart from ``white'' receiver
noise) in the timing residuals for one or more pulsars can be used to
limit the strength of the GW background in the Galaxy. Currently, the
best published limit, based on archival Arecibo data and Parkes timing
data for seven pulsars \cite{jhs+06}, is $h_c(f_{\rm 1 yr}) < 1.1\times
10^{-14}$. This limit does not significantly constrain current models
for super-massive black hole evolution and mergers in galaxies
\cite{jb03,wl03a,svc08} but does limit some models for the equation of
state of matter in the epoch of inflation \cite{gri05} and the tension in cosmic
strings \cite{dv05}.

\begin{figure}[ht]
  \includegraphics[height=0.4\textheight,angle=270]{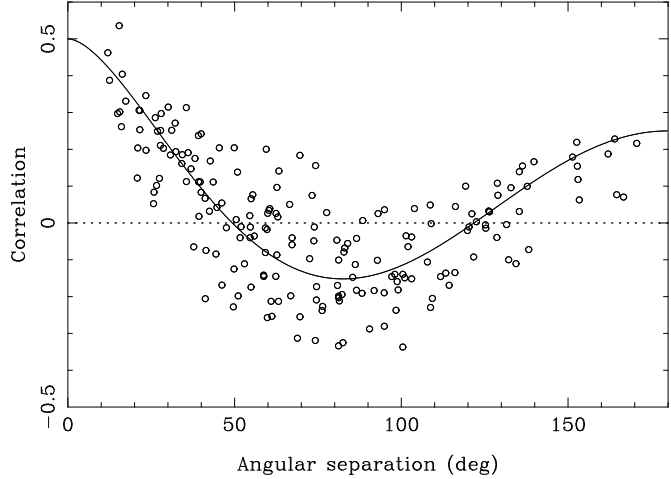}
  \caption{Correlation between timing residuals for pulsar pairs as a
    function of the angular separation of the two pulsars for an
    isotropic GW background in the Galaxy under general relativity
    \cite{hd83}. The computations are based on the PPTA sample
    of 20 pulsars with a strong GW signal that dominates the timing
    residuals \cite{hjl+09}.\label{fg:hd}}
\end{figure}

The principal aim of PTAs is not just to limit the stochastic GW
background in the Galaxy, but to {\em detect} it. This detection
depends on the correlated modulation of timing residuals for different
pulsars as a GW passes over the Earth. GW are quadrupolar in nature so
that modulations are opposite in sign for pulsars which are
$90^{\circ}$ apart on the sky. For an isotropic stochastic GW
background, the correlation between residuals in different pulsars
depends only on the angular separation of the pulsars. For general
relativity, the expected correlation is given by the ``Hellings and
Downs'' curve \cite{hd83} shown in Figure~\ref{fg:hd}. For pulsars
close together on the sky, the correlation is 0.5 rather than 1.0
because of the uncorrelated modulations due to GW passing over the
pulsars. These modulations also result in the ``self-noise'' apparent
in the scatter of points about the theoretical line; this noise is
independent of ToA precisions and limits the sensitivity of PTA
experiments in strong-field situations \cite{jhkm05,hjl+09}.

\section{Pulsar Timing Arrays and GW Applications}
Currently there are about 35 MSPs that are both sufficiently strong
and have sufficiently narrow pulse profile features to make them
useful for PTA projects, i.e., to give ToA precisions of $\lapp 1
\mu$s with observations times of an hour or so. Most of these are
being regularly timed by one or more of the three major PTA projects:
the European Pulsar Timing Array (EPTA), the North American
pulsar timing array (NANOGrav) and the Parkes Pulsar Timing
Array (PPTA). The EPTA combines data from the four large
radio telescopes in Europe: at Nan\c{c}ay, Effelsberg, Westerbork and
Jodrell Bank, with a fifth in Sardinia soon to be added
\cite{jsk+08a}. NANOGrav uses data from the Arecibo radio
telescope and the Green Bank Telescope \cite{jfl+09}. The PPTA
uses data from the Parkes 64-m radio telescope in Australia
\cite{hob05,man08,hbb+09}. A collaboration exists between these three
projects to form the International Pulsar Timing Array (IPTA)
\cite{haa+10}.

The PPTA project commenced in 2004 and regular timing of 20 MSPs
commenced in 2005 March. The project is a collaboration principally
between the groups at CSIRO Astronomy and Space Science (led by RNM
and G. Hobbs) and Swinburne University of Technology (led by M. Bailes
and W. van Straten) with major contributions from the University of
Texas at Brownsville (F. A. Jenet) and the University of California at
San Diego (W. A. Coles). Observations are made at intervals of 2 -- 3
weeks in three bands: 10~cm (3100~MHz), 20~cm (1400~MHz) and 50~cm
(700~MHz). The 10~cm and 50~cm observations use a dual-frequency
coaxial receiver while the 20~cm observations are generally made using
the centre beam of the Parkes 20-cm 13-beam receiver. A number of
back-end systems have been used since the project commenced: the
Wide-Band Correlator (maximum bandwidth 1024~MHz), a series of Parkes
Digital Filterbanks PDFB1 (256~MHz) and PDFB2, 3 and 4 (1024 MHz), the
Caltech-Parkes-Swinburne Recorder CPSR2 ($2\times$~64~MHz) and the
ATNF-Parkes-Swinburne Recorder APSR (1024 MHz). CPSR2 and APSR provide
coherent dedispersion whereas the other systems give up to 2048
frequency channels for off-line dedispersion.

Table~\ref{tb:ppta} lists the 20 MSPs observed as part of the PPTA
project. The timing results are from 2.3 -- 4.0 years of data recorded
with PDFB2 and PDFB4. Except for PSR J1045$-$4509 the data are
uncorrected for DM variations. In most cases, the observation time per
ToA is 64~min. Rms timing residuals after fitting for the pulsar
position, pulse frequency, its first time-derivative and the Keplerian
binary parameters (if applicable) and the band at which they were
obtained are listed. Most of the pulsars have rms timing residuals of
less than $1 \mu$s and four are less than 200 ns. We are therefore
approaching the level of timing precision needed for detection of the
expected GW signal in the Galaxy. Some improvement in these results
can be expected with improved calibration and signal processing
procedures.

\begin{table}[ht]
\caption{PPTA pulsars and their timing residuals}\label{tb:ppta}
\begin{tabular}{lccccc}
\hline
  \tablehead{1}{c}{b}{PSRJ}
  & \tablehead{1}{c}{b}{Pulse Period\\(ms)} 
  & \tablehead{1}{c}{b}{DM\\(cm$^{-3}$ pc)} 
  & \tablehead{1}{c}{b}{Orbital Period\\(d)} 
  & \tablehead{1}{c}{b}{Band} 
  & \tablehead{1}{c}{b}{RMS Residual\\($\mu$s)} \\
\hline
J0437$-$4715 & 5.757 & 2.65 & 5.74 & 10~cm & 0.055  \\
J0613$-$0200 & 3.062 & 38.78 & 1.20 & 20~cm & 0.72 \\
J0711$-$6830 & 5.491 & 18.41 & -- & 20~cm & 0.68 \\
J1022+1001 & 16.453 & 10.25 & 7.81 & 10~cm & 1.39 \\
J1024$-$0719 & 5.162 & 6.49 & -- & 20~cm & 0.84 \\
J1045$-$4509 & 7.474 & 58.15 & 4.08 & 20~cm & 2.60 \\
J1600$-$3053 & 3.598 & 52.19 & 14.34 & 20~cm & 0.49 \\
J1603$-$7202 & 14.842 & 38.05 & 6.31 & 20~cm & 0.46 \\
J1643$-$1224 & 4.622 & 62.41 & 147.02 & 20~cm & 0.80 \\
J1713+0747 & 4.570 & 15.99 & 67.83 & 10~cm & 0.23 \\
J1730$-$2304 & 8.123 & 9.61 & -- & 20~cm & 1.46 \\
J1732$-$5049 & 5.313 & 56.84 & 5.26 & 20~cm & 2.43 \\
J1744$-$1134 & 4.075 & 3.14 & -- & 20~cm & 0.18 \\
J1824$-$2452 & 3.054 & 119.86 & -- & 20~cm & 1.66 \\
J1857+0943 & 5.362 & 13.31 & 12.33 & 20~cm & 0.62 \\
J1909$-$3744 & 2.947 & 10.39 & 1.53 & 10~cm & 0.095 \\
J1939+2134 & 1.558 & 71.04 & -- & 10~cm & 0.18 \\
J2124$-$3358 & 4.931 & 4.62 & -- & 20~cm & 1.62 \\
J2129$-$5721 & 3.726 & 31.85 & 6.63 & 20~cm & 1.35 \\
J2145$-$0750 & 16.052 & 9.00 & 6.84 & 20~cm & 0.65  \\
\hline
\end{tabular}
\end{table}

\begin{figure}[ht]
  \includegraphics[height=0.32\textheight]{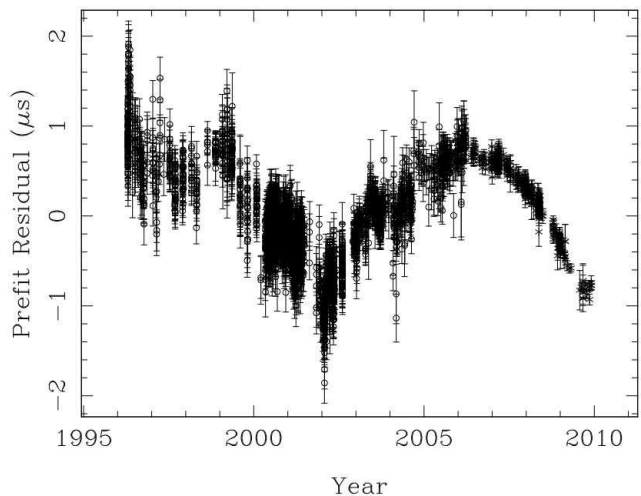}
  \caption{Timing residuals for PSR J0437$-$4715 for a 14-year data
    span formed by combining the Verbiest et al. \cite{vbv+08} data
    with more recent 10~cm PPTA data (from 2006.5). \label{fg:0437}}
\end{figure}

It is clear that longer data sets are needed to achieve the
sensitivity required to detect the GW background. Verbiest et
al. \cite{vbc+09} has shown that most of the PPTA pulsars have
sufficiently stable periodicities over a 10-year data span. However,
this analysis included arbitrary phase offsets between data obtained
with different instruments which absorbed some of the low-frequency
power in the residual spectrum. Figure~\ref{fg:0437} shows timing
residuals for PSR J0437$-$4715 over a nearly 14-year data
span. Offsets between the different instruments were measured using
local fits and then held fixed in the final fit which included just
the first period time derivative. These residuals obviously have red
spectrum, the origin of which is currently unknown. 

If such red noise is shown to be intrinsic to the pulsar and present
in most MSPs, it will have to be properly taken into account in
searches for GW and other analyses, for example, by using the Cholesky
technique \cite{chc+11}. It will not prevent detection of the GW
signal since, like the GW self-noise, it is uncorrelated between
different pulsars; it will however add noise to the correlations and
make the detection more difficult. The most effective way to overcome
such limitations is to increase the number of pulsars in the PTA
sample; this is a principal motivation for the IPTA collaboration. In
the future, the Square Kilometre Array (SKA) \cite{ckl+04} will have enough
sensitivity to obtain useful data from a much larger sample of
pulsars. This should allow not only the detection of GW signals
\cite{svc08,wjy+11} but also the study of the GW sources in some
detail. It will also allow detailed investigations of the properties
of the GW themselves, perhaps revealing non-Einsteinian behaviour
\cite{ljp08}. 

\begin{figure}[ht]
  \includegraphics[height=0.4\textheight,angle=270]{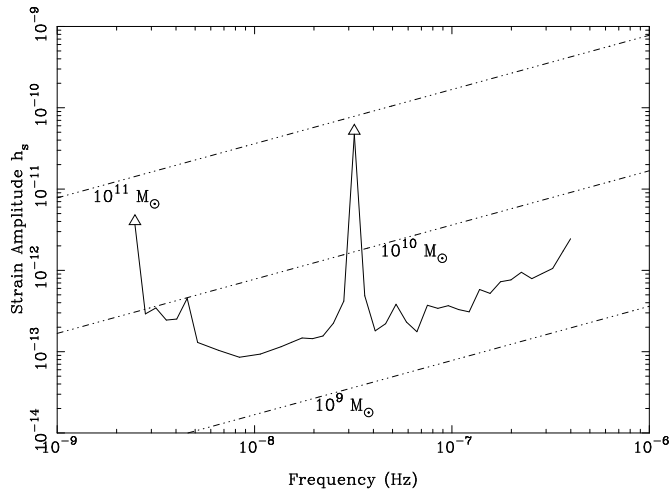}
  \caption{Sensitivity of a PTA based on the Verbiest et
    al. \cite{vbc+09} data set for detection of a source of GW in the
    Virgo cluster including all effects of the timing analysis
    \cite{yhj+10}. The dot-dashed lines show the expected signal for
    equal-mass binary black-hole systems in the cluster. \label{fg:sens}}
\end{figure}

While the stochastic background is predicted to be the strongest and
most easily detectable GW signal for PTAs, the SKA is likely to
provide sufficient sensitivity to detect and study individual sources
of GW \cite{svv09,yhj+10}. Figure~\ref{fg:sens} gives a realistic PTA
sensitivity curve applied to detection of individual GW sources in the
Virgo cluster \cite{yhj+10}, showing that a binary black-hole system
with component masses of a few $10^9$~M$_\odot$ and orbital period of
less than about 5 years would be detectable with present data
sets. Anholm et al. \cite{abc+09} showed that a sufficiently strong GW
source in the Southern Hemisphere can be localised to a few degrees
using the PPTA pulsar sample. A more uniform distribution of pulsars
on the sky, provided for example by the IPTA, would extend this
resolution to the Northern Hemisphere.

\section{Other PTA Applications}
PTA data sets have many applications besides the search for GWs. For
example, essentially all PTA data are recorded with full polarisation
information and hence can be used to study polarisation properties of
MSPs. This in turn has applications to studies of the pulse emission
process in pulsars and to studies of the Galactic magnetic
field. Timing data for individual MSPs can be used to study their
astrometric and binary properties, including proper
motions, parallaxes and studies of binary evolution and
neutron-star masses. Multi-frequency data sets can be used to
investigate interstellar dispersion and scattering, giving information
on the small-scale fluctuations in the interstellar medium \cite{yhc+07}. 

Pulsar timing analyses are dependent on a Solar-System ephemeris to
transfer observed ToAs to the Solar-System barycentre. Commonly used
ephemerides such as those from the Jet Propulsion Laboratory (e.g.,
\cite{sta98b}) are based on fits to large data sets including optical
astrometry and radar ranging of the planets. PTA observations give an
independent method of estimating the mass of the Solar-System
planetary systems. An error in an assumed planetary mass will induce a
signal in pulsar timing residuals at the orbital period of the
planet. This signal has a spatial dipolar signature as opposed to the
quadrupolar signature of GW and hence can be separated in PTA
data sets. Champion et al. \cite{chm+10} have used PPTA data together
with archival Arecibo data to measure the masses of the Solar-System
planetary systems with an uncertainty of order
$10^{-10}$~M$_\odot$. For the Jupiter system, this is a factor of four
better than the best result published in the open literature, but
still a factor of about 20 worse than the value obtained from
observations of the {\em Galileo} spacecraft. Longer and more complete
PTA data sets have the potential to give us the most precise masses
for some planetary systems.

\begin{figure}[ht]
  \includegraphics[height=0.4\textheight]{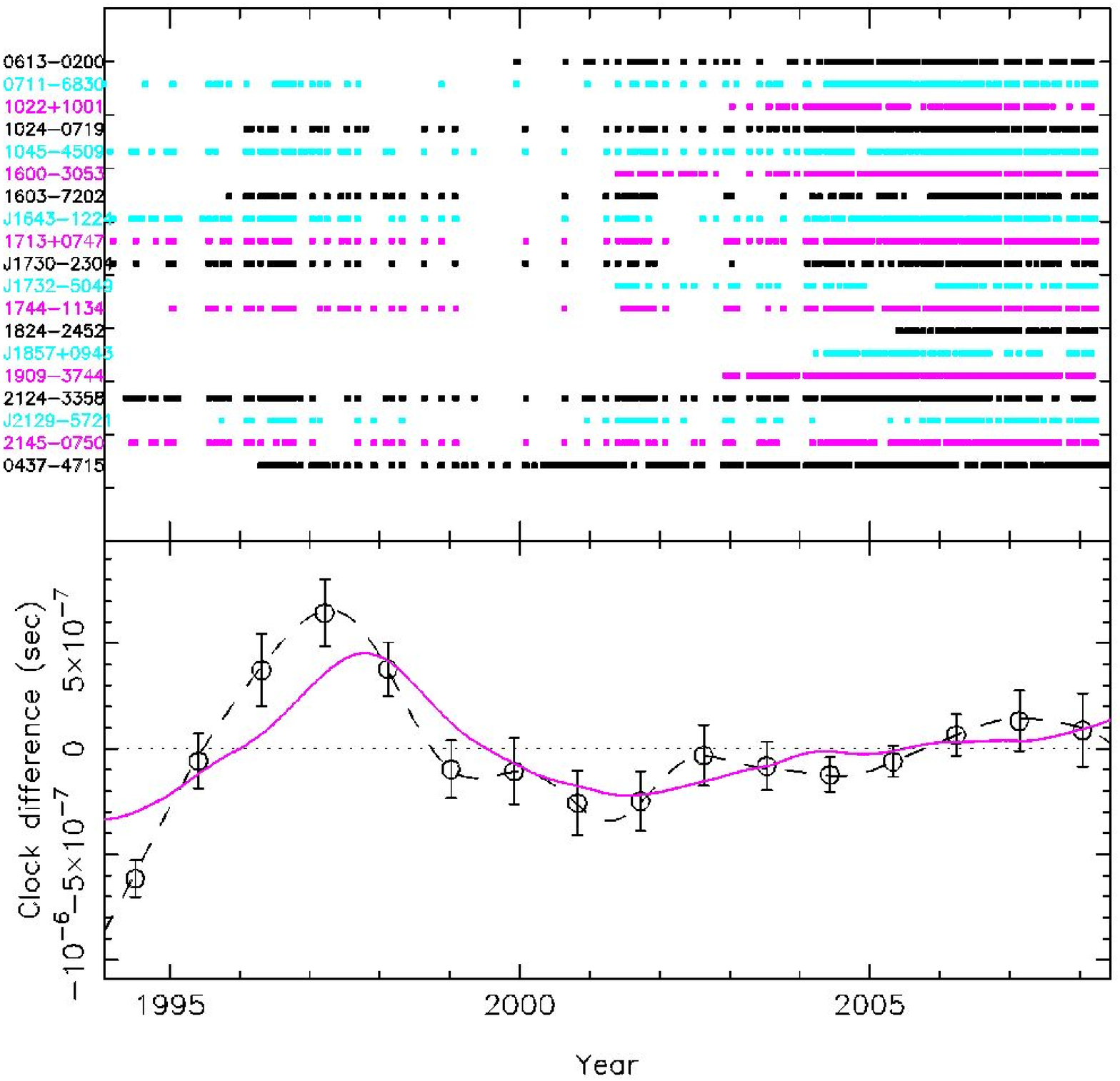}
  \caption{A pulsar timescale based on the extended PPTA data set
    \cite{hcmc10}. The upper panel shows the data
    sampling for the pulsars used in the analysis. In the lower panel
    the dashed line is the derived pulsar timescale relative to
    TT(TAI) (in the sense TT(TAI) $-$ TT(PSR)) and the solid line is
    TT(TAI) $-$ TT(BIPM2010) with a quadratic polynomial removed. \label{fg:clk}}
\end{figure}

PTA data sets also can be used to establish a standard of time which
independent of terrestrial atomic timescales \cite{pt96}. Fluctuations
in the atomic timescale introduce systematic residual fluctuations
which have a monopole signature, that is, they are the same for all
pulsars regardless of their position on the sky. Figure~\ref{fg:clk}
shows the results from an analysis of PPTA data by Hobbs et
al. \cite{hcmc10}. While there is good agreement between TT(TAI) and
the pulsar timescale TT(PSR) over the last five years or so, these
results indicate that TT(TAI) was running fast around 1997. The
retroactively revised atomic timescale TT(BIPM2010) shows
similar deviation at this time, confirming the accuracy of both the
pulsar results and the atomic timescale revision.

\section{Conclusions}
The realisation of Pulsar Timing Arrays is an exciting new development
in pulsar astrophysics. PTA data sets have many applications including
the detection of low-frequency gravitational waves and the
establishment of a pulsar-based standard of time. Combining of
existing and future data sets to form an International Pulsar Timing
Array will give improved results in all applications. Looking
further into the future, the greatly increased sensitivity provided by
the proposed Square Kilometer Array will make possible detailed
studies of phenomena presently near or below the level of
significant detection.

%%%%%%%%%%%%%%%%%%%%%%%%%%%%%%%%%%%%%%%%%%%%%%%%
%% BACKMATTER
%%%%%%%%%%%%%%%%%%%%%%%%%%%%%%%%%%%%%%%%%%%%%%%%

\begin{theacknowledgments}
  I thank my colleagues in the PPTA project and the staff of the
  Parkes Observatory for their efforts which have been vital to the
  realisation of the PPTA. The Parkes radio telescope is part of the
  Australia Telescope which is funded by the Commonwealth Government
  for operation as a National Facility managed by CSIRO.
 \end{theacknowledgments}




%\bibliography{journals,modrefs,psrrefs,crossrefs}

\end{document}